 \documentstyle[12pt,aaspp4]{article}

\begin{document}


\def\zz{\hang\noindent}

\def\kms{km s$^{-1}$}
\def\pix{pix$^{-1}$}
\def\deg{$^\circ$}

\def\mic{{$\mu$m}}

\def\h2o{H$_2$O}

\def\ak{{\it $A_K$}}

\def\teff{$T_{\rm eff}$}

\def\aple{$\mathrel{\hbox{\rlap{\hbox{\lower4pt\hbox{$\sim$}}}\hbox{$<$}}}$
}
\def\apge{$\mathrel{\hbox{\rlap{\hbox{\lower4pt\hbox{$\sim$}}}\hbox{$>$}}}$
}



\title{2 \mic \ Narrow--Band Imaging of the Sagittarius D H~II Region}

\author{R. D. Blum\altaffilmark{1}} 
\affil{Cerro Tololo Interamerican Observatory, Casilla 603, La Serena, Chile\\
rblum@noao.edu}

\author{A. Damineli\altaffilmark{2}}
\affil{JILA, University of Colorado\\Campus Box 440, Boulder, CO,
80309\\damineli@casa.colorado.edu}

\altaffiltext{1}{Hubble Fellow}
\altaffiltext{2}{Permanent address: IAG-USP, Av. Miguel Stefano 4200, 
04301-904, Sao Paulo, Brazil}


\begin{abstract}
We present 2 \mic \ narrow--band images of the core H~II region in the
Galactic star forming region Sagittarius D. The emission--line images
are centered on 2.17 \mic \ (Br$\gamma$) and 2.06 \mic \ (He~I). The H~II
region appears at the edge of a well defined dark cloud, and the
morphology suggests a blister geometry as pointed out 
in earlier radio continuum work. There is a deficit of
stars in general in front of the associated dark cloud
indicating the H~II region is located in-between the
Galactic center and the sun. The lesser spatial extent of the He~I
line emission relative to Br$\gamma$ places the effective
temperature of the ionizing radiation field below 40,000~K. The He~I 
2.06 \mic \ to Br$\gamma$ ratio and Br$\gamma$ / far infrared dust
emission put \teff \ at about 36,500~K to 40,000~K 
as derived from ionization models.

\end{abstract}


\newpage

\section{INTRODUCTION}

Sagittarius D (Sgr~D) is a well known star forming region seen 
toward the Galactic center (GC). The associated H~II region
(compact plus extended components) 
was noted in the survey of Downes et al. (1978)\markcite{dea78}. 
To be clear, the subject of this paper is the compact H~II region
(see the maps of Lizst 1992\markcite{l92}) which is superimposed
on a much larger, extended or 
diffuse H II region. These are very close to the prominent
super nova remnant (SNR) known as the Sgr~D SNR.
We will call the compact Sgr~D H II region ``H II region''
or ``Sgr~D H II region.'' We will call the larger region the ``extended
H II region.''
Sgr~D has been observed
at radio continuum wavelengths (Odenwald 1989\markcite{o89}; 
Lizst 1992\markcite{l92}; Mehringer et al. 1998\markcite{mea98}), 
in radio recombination lines (RRL, Whiteoak \& Gardner 1974\markcite{wg74},
Anantharamaiah \& Yusef--Zadeh 
1989\markcite{ayz89}), in CS and H$_{2}$CO lines 
(K\'{a}zes \& Aubry 1973\markcite{ka73}; 
Whiteoak \& Gardner 1974\markcite{wg74},
Lis 1991\markcite{l91}; Mehringer et al.1998\markcite{mea98}), 
and in the far infrared (Odenwald \& Fazio 1984 \markcite{of84}).

The location, and therefore the total luminosity and mass associated
with Sgr~D, is uncertain. Lis (1991)\markcite{l91} pointed out that
the line widths of the CS emission lines he associated with 
the H~II region are narrow and therefore not likely to
arise from a giant molecular cloud (GMC) in the GC nuclear disk as had
been previously thought. The line widths in the GC nuclear disk GMCs
are typically much larger than those observed toward Sgr~D or in the
GMCs of the Galactic disk. Lis (1991) suggested that Sgr~D could be
either in the foreground ($\sim$ 3.5 kpc) or part of the expanding
molecular ring, a complex located within a few hundred parsec of the
GC. Mehringer et al. (1998)\markcite{mea98} favor an in--or--beyond
the GC distance for Sgr~D (see the discussion below). 

The present images were obtained as part of an ongoing narrow--band
survey of the inner Galaxy designed to find emission--line stars.  The
2 \mic \ images provide the most detailed spatial and morphology
information to--date on the compact Sgr~D H II region and can be used
to infer information regarding the line of sight position of the Sgr~D 
H~II region and to constrain the properties of the stellar
source(s) responsible for the ionization.

\section{OBSERVATIONS AND DATA REDUCTION}

The narrow--band images were obtained on the nights 27 and 28 July
1996 using the Cerro Tololo Interamerican Observatory (CTIO) facility
infrared imager CIRIM on the 1.5--m telescope (f/8).  CIRIM employs a
NICMOS III array and is fully described in the CIRIM instrument manual
available from CTIO (http://www.ctio.noao.edu).
 
The images of Sgr~D were obtained as part of a narrow--band survey of
the inner Galaxy designed to detect emission--line stars (Blum \&
Damineli 1999) \markcite{bd99}. The narrow--band line filters employed
for this survey are 2.06 \mic \ (He~I, hereafter the 206 filter, $\sim$
0.5~$\%$), 2.08 \mic \ (C~IV, 208, $\sim$ 0.5$~\%$), and 2.17 \mic \ 
(Br$\gamma$, 217,
$\sim$ 1~$\%$). Continuum filters are also used so that line fluxes
can be measured; however,
due to large and non--uniform interstellar reddening, 
continuum filters are employed on each side of the lines to 
more accurately determine the continuum at the line position. The
continuum filters are 2.03 \mic \ (203, $\sim$ 0.5~$\%$), 2.14 \mic \
(214, $\sim$ 0.5~$\%$), and 2.248 \mic \ (225, $\sim$ 1~$\%$).  
The bandwidths are rough estimates from transmission scans provided
by the manufacturers, Omega Optical (203, 206, 208, 214) and
Barr Assoc. (217, 225). 
The 206 filter is slightly more peaked than the 217 filter. 
We estimate that the 206 line fluxes will have an uncertainty \aple 10
$\%$ due to transmission profile.

The 208 filter is designed to detect ionized carbon (C~IV) 
in emission--line stars and is
essentially another continuum filter for purposes of investigating H~II 
regions. The 225 filter is a molecular hydrogen filter, but is used as a 
continuum point for the emission--line star survey where such emission 
is expected to be absent. 
In the present case, there could be some H$_2$
emission contributing to the 225 filter, but we expect it to 
be small relative to H~II region emission as is the case,
for example, in the Orion nebula (Bautista et al. 1995\markcite{bpd95}). 
Furthermore, any contribution by H$_2$ is mitigated
by the fact that we use a linear 
fit to the four continuum points (203, 208, 214, 225)
to determine the continuum at the emission--line positions. The derived
continuum at 2.248 \mic \ is \aple 10 $\%$ higher than
the derived continuum at 2.06 \mic \ for 5$''$ to 55$''$ radius 
apertures centered on the H~II region.

The plate scale of CIRIM at f/8 is approximately
1.16$''$ \pix, 1.17$''$ \pix, 1.18$''$ \pix, 1.18$''$ \pix,
1.19$''$ \pix, and 1.19$''$ \pix \ for the 225, 217, 214, 208, 206, and
203 filters, respectively.
The images were taken in strips one degree long in right ascension (and
one frame high, or five arcmin in declination), each image
offset by 99$''$ from the previous. The strips were obtained by
stepping the telescope in one direction for one full degree, 
then stepping back in the reverse direction in the 
next filter. 
When a complete strip was completed in all filters, an offset north was
made and a new strip begun. The northern offset was made
with an eastern one to keep the survey area centered along the Galactic
plane. A total of nearly three square degrees has been observed
to date. 

Individual images had exposure times of 5 seconds in the 225 and 217 filters
and 15 seconds for the remaining four filters.
The seeing on individual frames was about 1.9$''$ to 2.0$''$ (measured
FWHM). This image quality 
was degraded by $\sim$ $5-6\%$ after combining the images (see below).

All basic data reduction was accomplished using
IRAF\setcounter{footnote}{2}\footnote{IRAF is distributed by the
National Optical Astronomy Observatories.}. Each image
was flat--fielded using dome flats and then sky subtracted using
a median combined image of approximately 35 data frames. 
A set of seven sky subtracted frames covering the area around the 
H~II region (in each filter) was combined to form a final image in 
each filter. Each of these six final images was then
trimmed on either end 
such that it comprised an area everywhere
sampled by three individual frames and was 
roughly centered on the H~II region. This trimming of the
ends of the combined images resulted in a final area of 
$\sim$ 8.1$' \times 5.0'$ for each image. No pixel shifts were
made or trimming
done in the declination direction to obtain the final images.

Each narrow--band image was flux calibrated using the the Elias et al.
(1982)\markcite{eea82} standard HD~161743 (B9).  This star was
observed three times during the night of 27 July over a range in
airmass bracketing the observations. Only the 214 filter was observed
on 28 July and the standard star observations were obtained at an
airmass within a few percent of the Sgr~D 214 data. The standard
deviation in the mean of the standard star measurements was less than
one percent for each of the six filters.  We adopt the least square
fit uncertainty for each filter as a function of airmass as the
uncertainty for the flux calibration (except for 214). This was less
than five percent in all cases and typically a few percent.  The flux
for each filter was set by the ratio of the 2.2 \mic \ flux to the
flux at the filter wavelength as determined from a blackbody with \teff
\ of 10,600~K, appropriate for a B9 V (Johnson 1966\markcite{j66}).

HD~161743 is a B9~IV star and thus has intrinsic Br$\gamma$ absorption.
We have made a small correction (0.10 mag) to the 217 magnitudes to
account for this. Measurements of the G5~V star HD~1274 (Carter \&
Meadows 1995\markcite{cm95}; transformed to the CTIO/CIT system, Carter
1990\markcite{c92})\markcite{c90} suggest an 0.05 mag difference in
the Br$\gamma$ absorption relative to HD~161743.  The Ali et
al. (1995)\markcite{ali95} and Hanson, Conti, \& Rieke (1996\markcite{hcr96})
spectroscopic observations of B and G stars
suggest G5~V stars have non-negligible Br$\gamma$, about half that
expected for a B9~IV star, hence the correction of 0.10 mag.

\begin{figure}
\plotfiddle{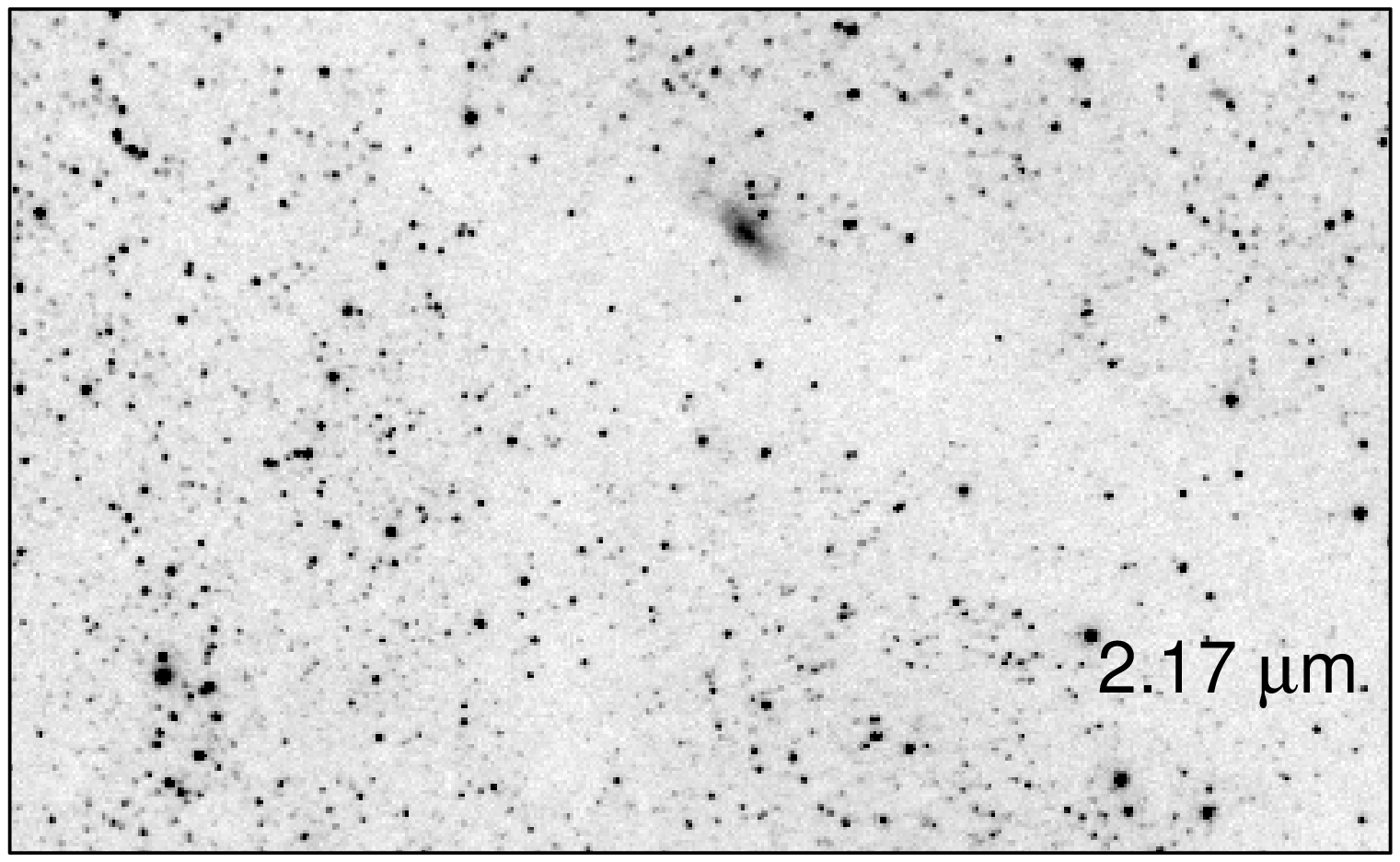}{6. in}{0}{60}{60}{-200}{100} 
\plotfiddle{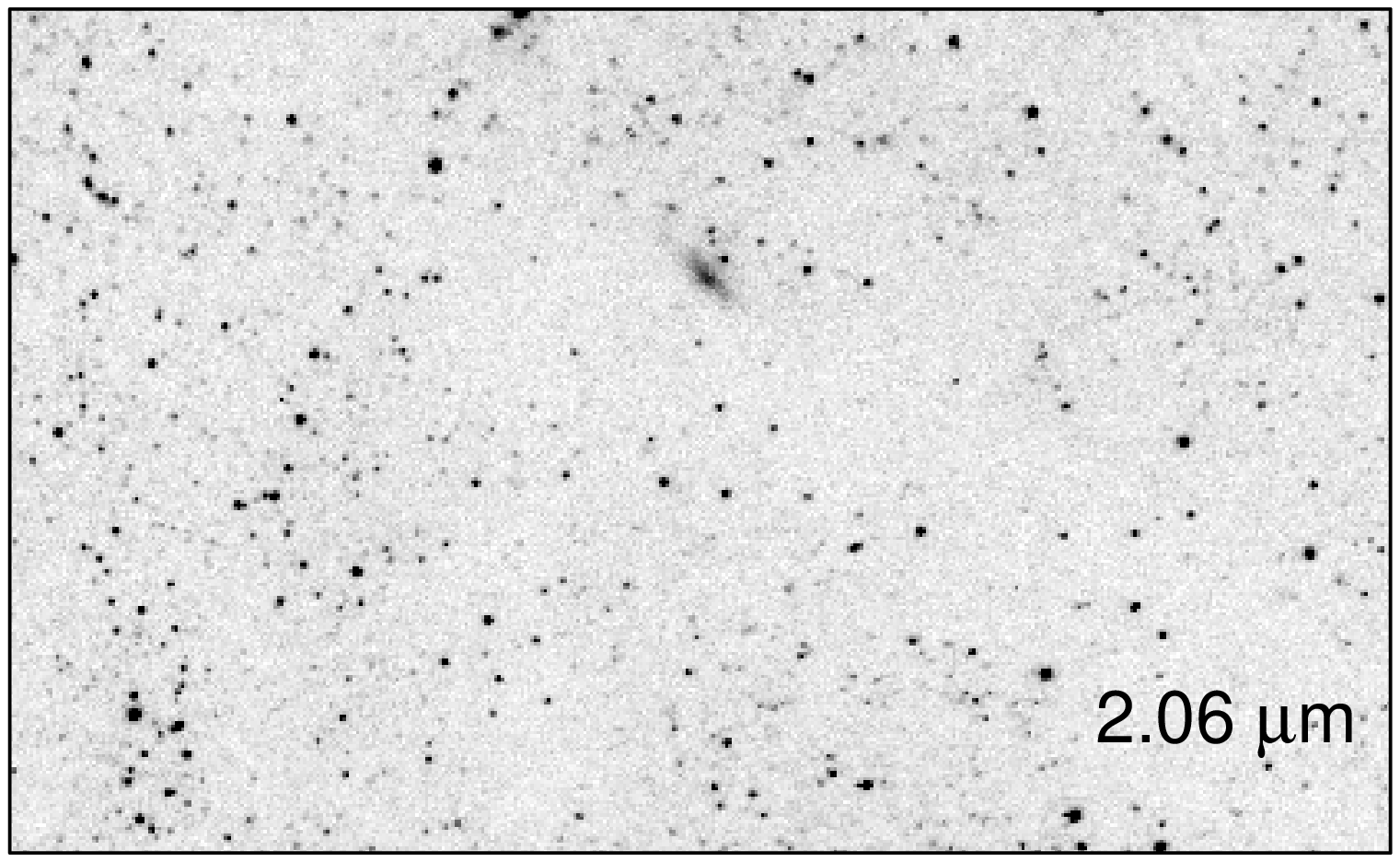}{.1 in}{0}{60}{60}{-200}{-35} 
\plotfiddle{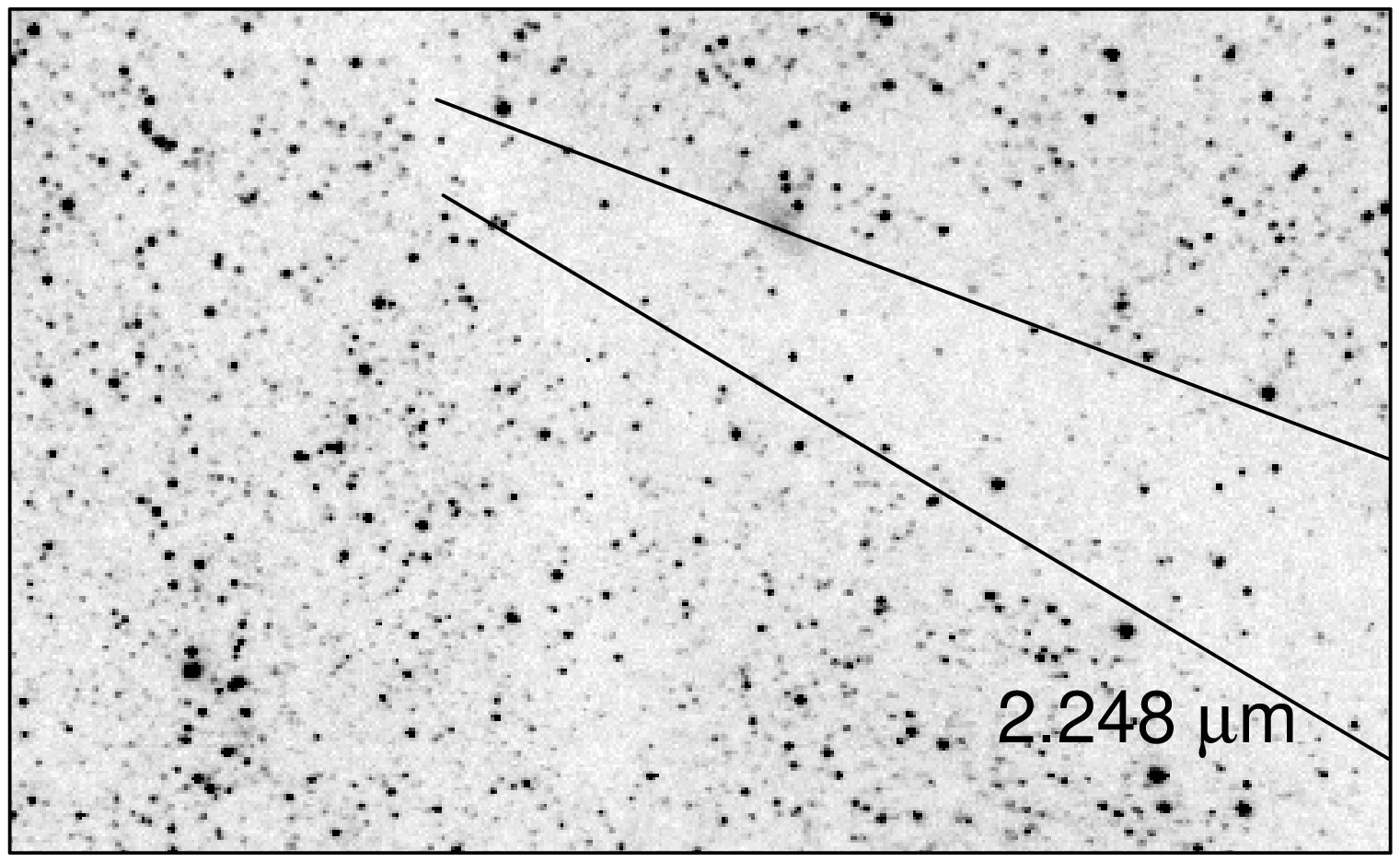}{.1 in}{0}{60}{60}{-200}{-170} 
\figcaption[]{
Narrow--band images. North is up and East to the left. 
The scale is 1.2$''$ \pix \ 
and each image covers roughly 8$'$ $\times$ 5$'$. The images are 
2.17 \mic \ (Br$\gamma$ $+$ continuum),
2.06 \mic \ (He~I $+$ continuum), 
and 2.248 \mic \ continuum from top to bottom respectively.
The lack of stars in projection against the dark cloud
(indicated by by the lines 
in the 2.248 \mic \ image), suggests it is relatively in the 
foreground; see text. Not shown are the 203, 208, and 214 images.
Continuum subtracted images are shown in Figure~\ref{images_sub}.
\label{images}
}
\end{figure}

Aperture photometry was computed in all six filters centered on the
compact H~II region using the IRAF ``PHOT'' package. The 
Br$\gamma$ peak defines the aperture center on all images. 
This peak is the same in Figure~\ref{images}b and Figure~\ref{images_sub}a
to about 1 pixel.
A correction was made to the background by specifying a background
aperture located on the darkest area of the cloud about one arcminute
to the SW from the edge of the H~II region. This background eliminates
incomplete sky subtraction resulting from differences in the combined
image sky frame and the subset of images for the H~II region, as well
as a low level background due to the extended H~II region.  The
largest residual sky was three percent for the 217 image.  Of course,
nonuniform emission from the extended H~II region will not be
accounted for.

In addition to aperture photometry, DoPHOT (Schecter, Mateo, \& Saha
1993)\markcite{sea93}  point--spread--function (psf) fitting
photometry was obtained for point sources in the field. Version 3.0 of
the code was used. The point source photometry is primarily for use in
the search for emission--line stars but is useful in the present case
to estimate interstellar reddening to stars near the H~II region. For
the psf fitting photometry, an additional aperture correction was
determined for each image to transform the DoPHOT magnitudes to the
standard star magnitudes.  This was accomplished by measuring aperture
magnitudes for 10--15 stars on each image. The corrections obtained in
this way had standard deviations in the mean of two to four percent.

\section{RESULTS AND DISCUSSION}

The narrow--band images in the 206, 217, and 225 filters
are shown in Figure~\ref{images}; the 203, 208, and 214 filter images
are not shown.  The compact
H~II Region is the obvious feature toward the center--north of these
images. Continuum subtracted line images are shown in Figure~\ref{images_sub}.
The continuum subtracted images use the 225 and 214 images for the
Br$\gamma$ and He~I 2.06 \mic \ continuum, respectively.
The extent in the Br$\gamma$ image (Figure~\ref{images_sub})
is roughly 21.5$''$ $\times$ 8.5$''$ FWHM, 
compared to the 25$''$ $\times$ 10$''$ in the 1616 MHz VLA maps
reported by Liszt (1992\markcite{l92}) and 20$''$ $\times$ 10$''$ 
reported by Mehringer et al. (1998\markcite{mea98}).
Mehringer et al. (1998\markcite{mea98}) analized the 1616 MHz (18 cm) data
of Liszt (1992\markcite{l92}) and also new 6 cm data.
The measured FWHM of the He~I image (Figure~\ref{images_sub}) is 
18.8$''$ $\times$ 6.3$''$. We verified the source position by matching
point sources from a visual image centered on the radio position of
Liszt\markcite{l92} (1992, source 3, $\alpha_{1950}, \delta_{1950} 
=$ 17$^{\rm h}$ 45$^{\rm m}$ 32$^{\rm s}$.1, $-$28\deg \ 00$'$ 43$''$) 
to the infrared
sources on our images. The visual image was obtained from the
digitized sky survey (DSS\footnote{The Digitized Sky Surveys were
produced at the Space Telescope Science Institute under
U.S. Government grant NAG W-2166.  Based on photographic data obtained
using The UK Schmidt Telescope.  The UK Schmidt Telescope was operated
by the Royal Observatory Edinburgh, with funding from the UK Science
and Engineering Research Council, until 1988 June, and thereafter by
the Anglo-Australian Observatory.  Original plate material is
copyright (c) the Royal Observatory Edinburgh and the Anglo-Australian
Observatory.  The plates were processed into the present compressed
digital form with their permission.}).  No indication of the H~II
region appears on the DSS image.

\subsection{Image Morphology and Reddening to the H~II Region}

The morphology of the H~II region appears to be one of a blister on
the edge of a clearly delineated dark cloud as pointed out by Odenwald
(1989) from radio continuum images. The very sharp edge to the
line emission toward the cloud leaves little doubt
that the cloud and H~II region are physically related.
Figure~\ref{cont} shows contour plots of the 225 (2.248 \mic \
continuum) filter and Br$\gamma$ filter images.  The H~II region is
much less extended in the continuum, and there is a strong suggestion
of three embedded sources at the location of the peak Br$\gamma$
emission. None of these appears truly point--like but this may be
largely due to their close proximity to one another and the 
contamination by extended emission. Alternately, each peak may be a group
of embedded stars or perhaps a knot of gas.

In stark contrast to the obvious dark cloud is the crowded stellar
field, well known at 2 \mic \ toward the GC. The lack of stars
in projection against the cloud strongly suggests it is more or less in the
foreground. Indeed, the stellar density peaks so dramatically 
toward the inner Galaxy (see Kent 1992\markcite{k92} and references therein) 
that it is difficult to construct a plausible model
which does not place the dark cloud between the sun and the GC.

\subsubsection{Interstellar Extinction to Individual Stars}

We have determined
$A_K$ for individual stars with measurements in all six narrow--band
filters by fitting to three model flux distributions, a 10,000~K
blackbody, a 5,000~K blackbody, and a bulge late M giant
(none of these stars had narrow--band indices suggesting an
emission--line object). The latter
model was constructed from spectra of bulge M giants (Terndrup,
Frogel, \& Whitford 1991\markcite{tfw91}) as tabulated by Blum,
Sellgren, \& DePoy (1996$b$)\markcite{bsd96}. The primary difference
between models is the intrinsic absorption by H$_2$O in the late M
giants toward the blue end of the $K-$band which mimics the effect of
reddening. There is more curvature in the M giant spectra combined
with reddening than for a reddened blackbody, and so the former model
can improve the fit substantially for some stars. Our filters are
all blueward of the CO bandhead at 2.3 \mic \ which is prominent in M giants. 
A Mathis (1990)\markcite{m90} reddening law was adopted and $A_K$ values
assigned to the individual stars based on the best fit among the three
models (typically the spread in $A_K$ is about
1.5 mag for the 10,000~K and M giant models).

For 205 stars with six filter photometry, we find $A_K$ $=$ 2.3 $\pm$ 
1.9 mag. Concentrating on
the area around the H~II region (to be quantitative we will consider a
3.5$'$ $\times$ 2.5$'$ area centered on the H~II region with roughly
equal area on and off the dark cloud), we find the
majority of stars have $A_K$ $\leq$ 1.5 mag (18 of 34 stars with measured
$A_K$ in this sub--region). 
Some stars were too faint or in crowded regions and thus did not
have photometry in all filters. However, in this region, all but one 
of the stars seen in projection against
the dark cloud can be accounted for either because 
they have a measured $A_K$ value, or because they appear on the DSS image.
The one star with no DSS counterpart and no $A_K$ value 
is $\sim$ 1$'$ away from the H~II region. Another 
star which appears near the middle of the dark cloud has $A_K$ $=$ 5.5 mag.
This may be an embedded object. Except for these two objects,
the remainder of the stars seen
in projection against the cloud have $A_K$ \aple 1.5 mag
or DSS counterparts.
We have obtained estimates of the $V$ magnitudes for the
latter (seven stars) 
from the DSS image and revised photometric calibration$^4$ available
from the Catalog and Surveys Branch on the Space Telescope Science Institute
www pages. These stars have $V$ \apge 15.5 mag. The majority are upper limits
to the brightness because they are fainter than the faint limit of the
non--linear 
calibration, but within 40 $\%$ of the counts of the faint end of the 
calibration ($V =$ 15.91 mag $\pm$ 0.5 mag).
We derived estimates of the corresponding $K$ magnitudes 
for each of these stars from aperture photometry on the
225 image ($V - K$ from 1.8 to \apge 3.8). 
All are consistent with late type (K or M) foreground dwarfs.

The closest star to the H~II region (\aple 8$''$)
with no DSS counterpart has $A_K$ $=$ 1.0 mag (star ``A'' in Figure~\ref{cont}).
This star had a best fit with the M giant model (reduced $\chi$ square $<$ 1). 
The 10,000~K blackbody model
(reduced $\chi$ square = 1.5) gives $A_K$ $=$ 2.4 mag.
Of course, off the dark cloud but within the near region, there are
stars with larger $A_K$. This includes
two located 10--15 arcsec north of the center of the
H~II region (stars ``B'' and ``C'' in
Figure~\ref{cont}) with indicated $A_K$ $\sim$ 5 mag. 
Their proximity to the H~II region suggests
they may be young stellar objects with infrared excesses. Alternately,
they may be background 
sources seen through the large column of material 
at the edge of the cloud.

\subsubsection{Star Counts}

Next, we considered star count
models in order to compare  
the expected numbers of stars in the field relative to those in projection
against the cloud with what is observed in the images.
Our models have
two extinction components, a uniform screen and the cloud
itself. The uniform screen affects all stars in the field of view which 
lie behind it. The cloud can be thought of as a source of infinite 
extinction so that we see only foreground stars in front of it. This will
be approximately true toward the center regions of the cloud and less so
near the edges where background stars will be seen. In addition, embedded
sources in the cloud could be projected upon the cloud at any location.
Embedded sources should be quite reddened. Foreground stars projected
against the cloud (or in the rest of the field) 
should be relatively blue. 
The position of the cloud 
along the line of sight will set (roughly) the relative numbers of stars
in the cloud foreground compared to the background 
field. The H~II region appears clearly
related to the dark cloud; thus, any constraints on the location of the dark 
cloud should apply to the H~II region.

We have detected stars in the narrow--band images down to $K \approx 14.5$
mag. A histogram of the 225 filter narrow--band magnitude
($K_{225}$, $\lambda$ $=$ 2.248 \mic) is shown in Figure~\ref{225}.
There is a break at about $K_{225}$ $=$ 11 mag which suggests the 
completeness limit (likely set by crowding) is at this 
magnitude or brighter. We estimated the actual completeness limit by adding
artificial stars to the $K_{225}$ image. Extracting stars from ten artificial 
images (with 300 artificial stars added to each), we find the $K_{225}$ counts
are about 95~$\%$ complete to $K_{225}$ $=$ 11th magnitude, 
87~$\%$ to 12th magnitude, and 77~$\%$ to 13th magnitude. 

The cumulative counts in the 225
filter are 115 to $K_{225}$ $\leq$ 11 mag.
The number with $K_{225}$ $\leq$ 11 mag and
in projection against the dark cloud is seven for the darkest
central strip which cuts from the north east to the south west
on the frame (see Figure~\ref{images}c). Accounting
for an area of 18~$\%$ of the frame covered by this darkest region of the
cloud, we find there are \aple 29~$\%$~$\pm$~10~$\%$ 
as many stars in the foreground of the
dark cloud per unit area as there are in the remainder of the field
(The uncertainty assumes Poisson statistics in the star counts.).
We can use the models to estimate the distance where this would occur.

We have computed cumulative star count models as described by Blum et
al. (1994\markcite{bea94}). These models are based on the Kent
(1992\markcite{k92}) two component bulge and disk model
and assume the sun--to--GC distance is 8 kpc. 
The associated luminosity functions are for a broad band $K$ filter
($\lambda$ $=$ 2.2 \mic).
We will assume the 225 filter counts ($\lambda$ $=$ 2.248 \mic) are  
equivalent to those in $K$. 

A model with a uniform screen of extinction ($A_K = 2.2$ mag at 4 kpc from
the sun), matches the actual counts: 121 stars
with $K$ $\leq$ 11 mag (correcting for 5~$\%$ incompleteness).
This model is not too
unreasonable given that the neutral hydrogen in the disk peaks rather
strongly at this
distance from the sun (for a sun--to--GC distance of 8 kpc,
Burton 1988)\markcite{b88}, and the average $A_K$ as determined from individual
stars is 2.7 mag $\pm$ 2 mag (for stars with $K_{225}$ $\leq$ 11.0 mag). 
This $A_K$ value should be viewed
with caution since only stars with measurements in all six filters were
used (64 stars of the 115 with $K_{225}$ $\leq$ 11.0 mag). 
Still, it is in agreement with 
mean values to other fields toward the inner Galaxy (Catchpole, Whitelock,
\& Glass 1990\markcite{cwg90}; Blum, DePoy, \& Sellgren 1996$a$
\markcite{bds96}). 

If the dark cloud
blocked all star light behind it and the 29~$\%$ of the stars we see
are foreground (indeed, many of the infrared sources in projection against the 
cloud are seen in the DSS image), then the model would predict its
distance at 6.0 kpc $\pm ^{1.5} _{2.5}$ kpc.  
The {\it screen} of extinction may be placed variously between 2 and 6.2 kpc,
and the extinction varied between 2.0 and 3.0 mag to still match
the observed counts. For this range of parameter space,
the cloud can be located between 3.9 and 7.9 kpc. 
It is possible to have models with the dark cloud quite
close to the GC because the bulge stellar
distribution is very peaked. Therefore, models with the screen 
nearer the sun will have fewer nearby disk stars, and so,
to reach 29~$\%$ of the total
requires the dark cloud to be closer to the GC. However, because the 
bulge distribution is so peaked, it cuts off very rapidly after the GC. For
all the models (including a model with no extinction), even the 
point at which 50~$\%$ of the total counts (to $K$ $=$ 11 mag)
are reached is always on the near side of the GC. 

\subsection{Recombination Lines}

\begin{figure}
\plotfiddle{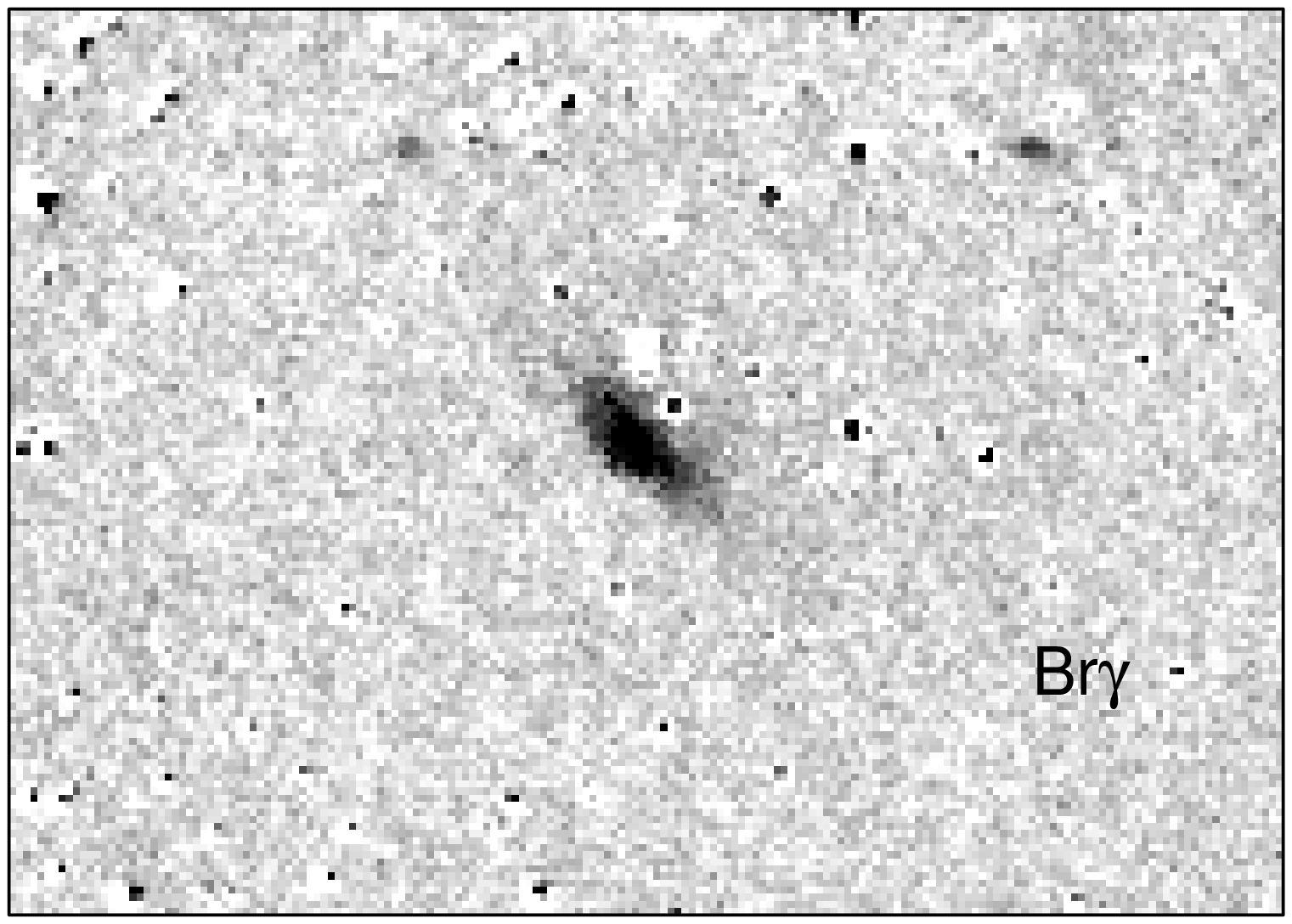}{5. in}{0}{60}{60}{-200}{50}
\plotfiddle{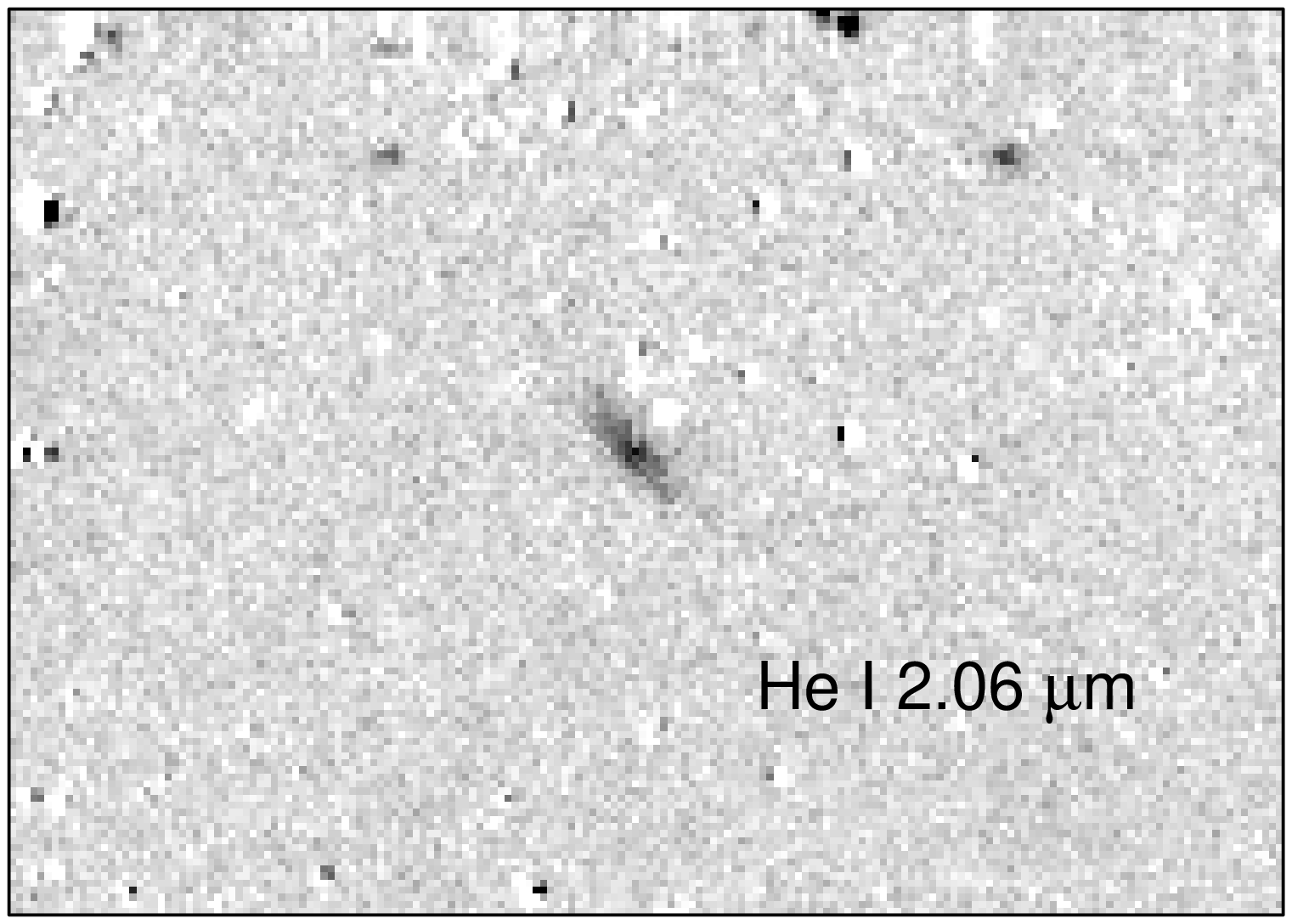}{.1 in}{0}{60}{60}{-200}{-110}
\figcaption[]{
Continuum subtracted line images. 
North is up and East to the left.
Each image covers roughly 3.5$'$ $\times$ 2.5$'$ centered on the 
H~II region. The images are Br$\gamma$ (217$-$225; see text),
and He~I 2.06 \mic \ (206$-$214).
The sharp edge to the Br$\gamma$ and He~I 2.06 \mic \ emission
toward the dark cloud indicates that the H~II region is physically associated
with it. The appearance and measured extent of the Br$\gamma$ emission
is larger than that for the He~I emission;
this is confirmed in Figure~\ref{faper}.
\label{images_sub}
}
\end{figure}

\begin{figure}
\plotfiddle{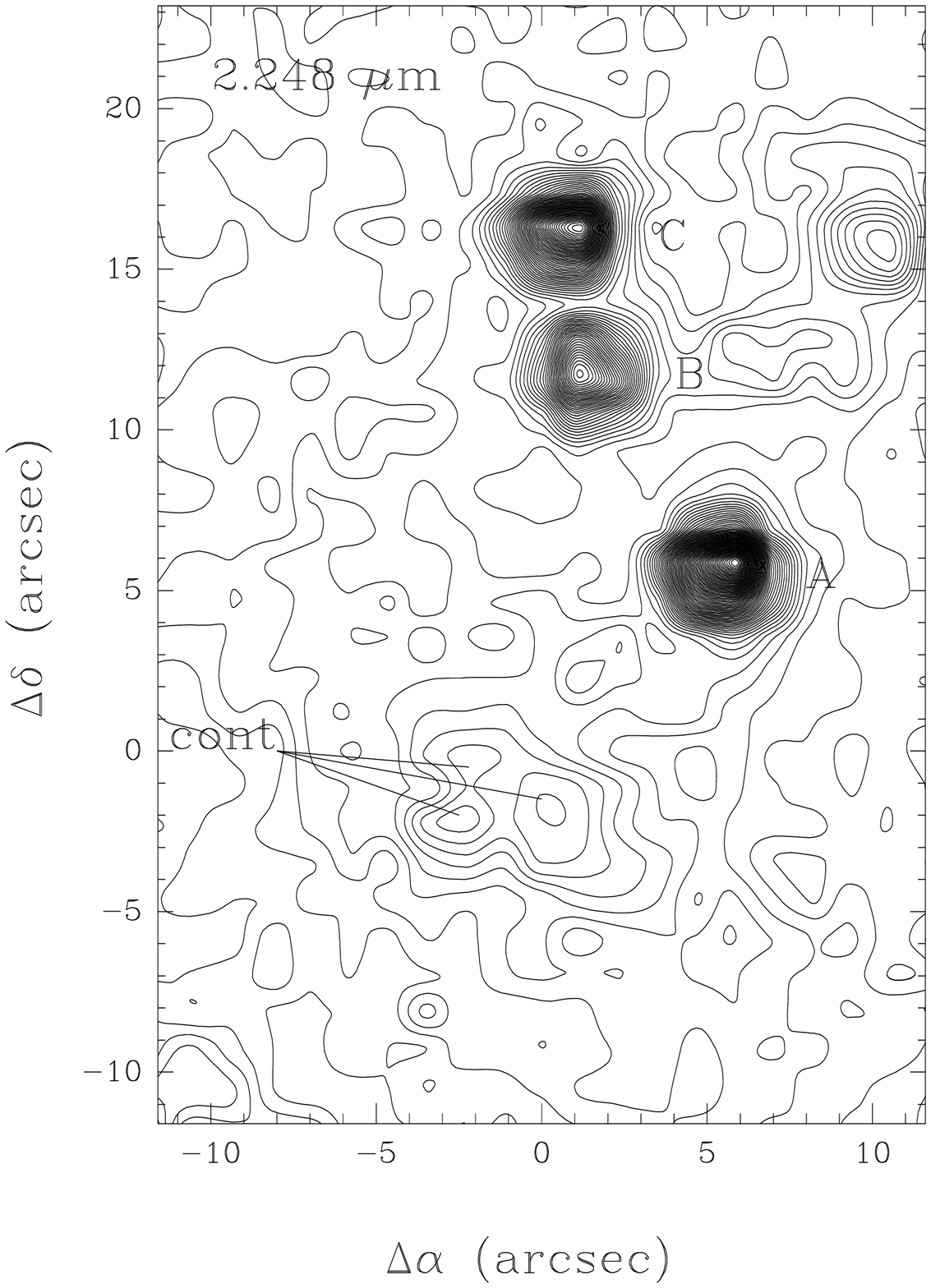}{5 in}{0}{55}{55}{-275}{-50}
\plotfiddle{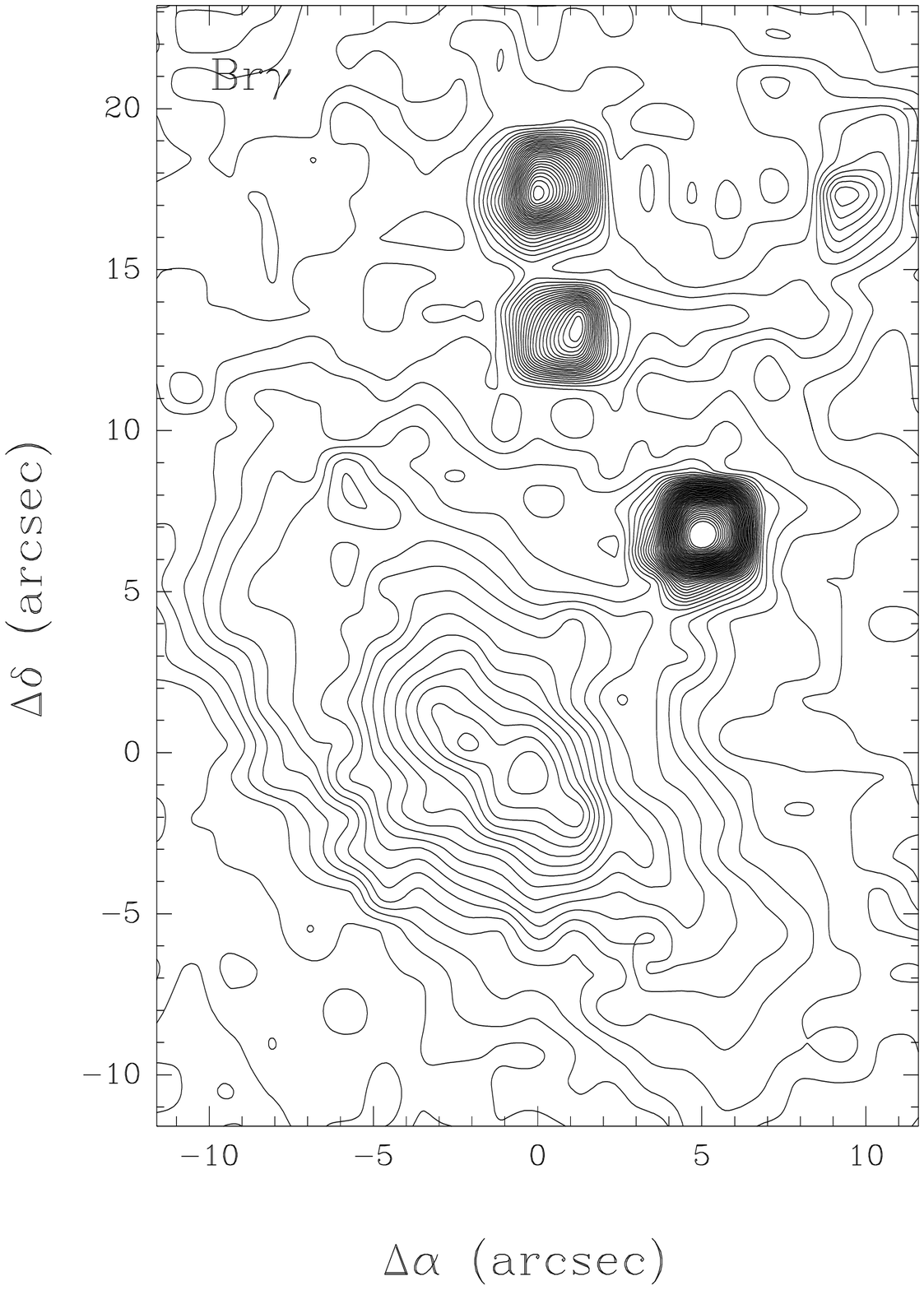}{.01 in}{0}{55}{55}{-35 }{-18}
\figcaption[]{
Contour plot of the region surrounding the H~II region for the 2.248 \mic \
continuum (left) and 2.17 \mic \ (right) images (Figure~\ref{images}). 
The contour interval is 1.7$\times$10$^{-15}$ erg cm$^{-2}$ s$^{-1}$ 
arcsec$^{-2}$
and 2.0$\times$10$^{-15}$ erg cm$^{-2}$ s$^{-1}$ arcsec$^{-2}$
in the 2.248 \mic \ and 2.17 \mic \ plots, respectively.
The measured FWHM of point sources 
is approximately 2.1$''$
and 2.0$''$ for the 2.248 \mic \ and 2.17 \mic \ images, respectively.
The continuum plot shows three
somewhat extended objects (indicated by ``cont'') 
which coincide with the peak Br$\gamma$ emission.
The impression is one of embedded sources, one or more of which likely
ionize the H~II region. We can not rule out the possibility that any or  
all of these sources is a knot of gas or group of stars.
Source ``A'' is possibly a nearby but foreground star to the H~II region;
see text.
Sources ``B'' and ``C'' are potentially young stellar objects associated
with the H~II region and dark cloud by virtue of their very red color.
Alternately, ``B'' and ``C'' may be background sources 
viewed through the large column of material at the
edge of the cloud.
\label{cont}
}
\end{figure}

\begin{figure}
\plotone{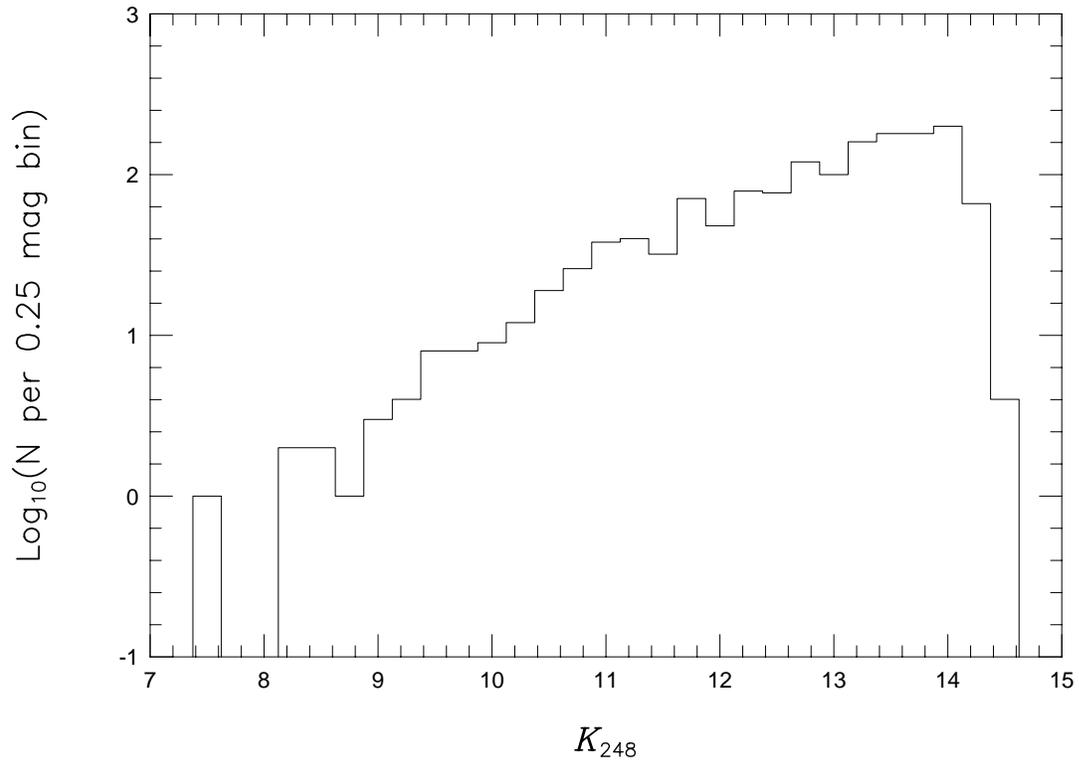}
\figcaption[]{
$K_{225}$ observed luminosity function. Number of stars detected at three sigma
in the 2.248 \mic \ image as a function of magnitude. The 225 filter 
is the deepest of the narrow--band images presented here. The 
star counts appear to rise less steeply at 
$K_{225}$ \apge 11 mag, which suggests the completeness limit 
(likely due to crowding) is \aple 11 mag. This is verified by artificial
star experiments; see text. 
\label{225}
}
\end{figure}

\begin{figure}
\plotone{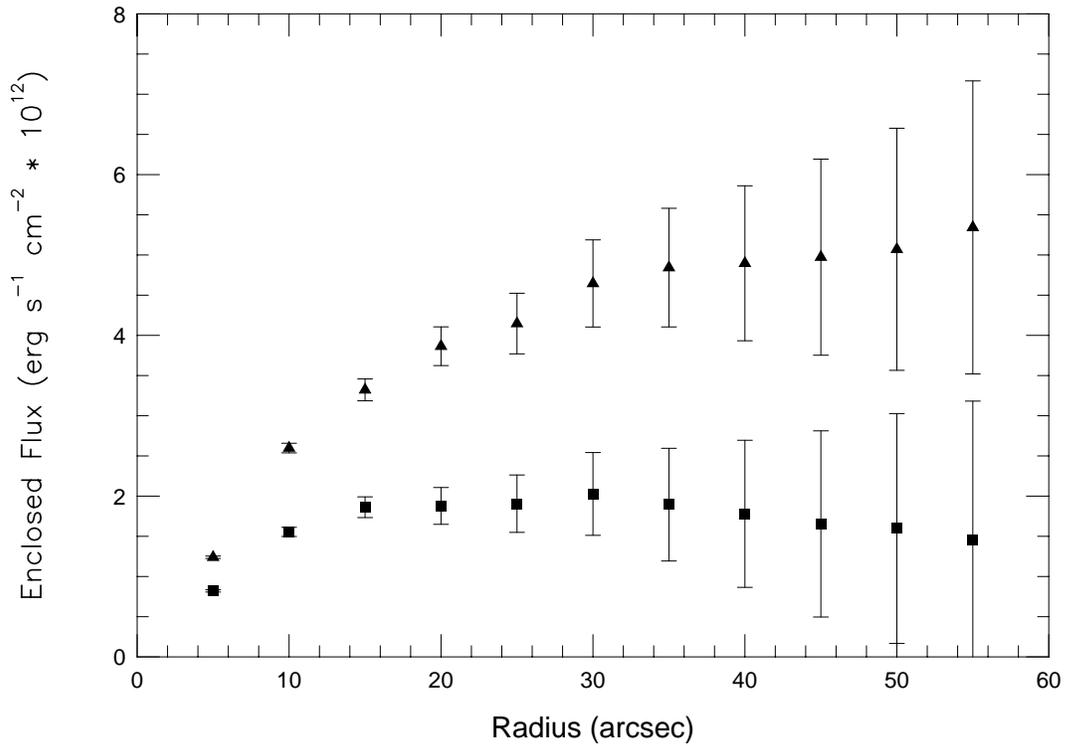}
\figcaption[]{Enclosed flux vs. radius for 
the Br$\gamma$ ({\it filled 
triangles}) and He~I 2.06 \mic \ ({\it filled squares}) emission lines. 
Both curves have sharp changes in slope which we take to indicate the angular 
extent of the emission. The
smaller extent of the He~I emission suggests a relatively cool ionizing 
continuum. The slight rise in flux beyond 35$''$ in the Br$\gamma$
image is possibly due to the associated extended H~II region; see text.
\label{faper}
}
\end{figure}

The integrated flux as a function of aperture radius is shown 
for the He~I 2.06 \mic \ line and Br$\gamma$ in Figure~\ref{faper}
where the center of the apertures was taken as the peak in the
Br$\gamma$ image (Figure~\ref{images_sub}a).
The line fluxes have been continuum subtracted. The continuum
at the line center was taken from a linear 
least--squares fit to the 203, 208, 214, and 225 fluxes.
Each curve reaches a
point where the increase in flux between successive apertures changes sharply.
For the He line, a clear maximum is reached near 15$''$.
The apparent decrease in flux in He~I 2.06 \mic \ at larger radii
may be due to a slight over--subtraction of the background. For Br$\gamma$, the
slope 
change occurs at about 35$''$ where the total flux remains nearly constant. 
The possible increase at large radii 
may be due to non--uniform emission in the extended H~II region. 
The He~I and Br$\gamma$ radial extents are consistent with the FWHM values
reported above; there is little additional flux beyond a 
radius of $\sim$ 20$''$--25$''$ (see Figure~\ref{faper}). 
As suggested by the appearance of the H~II region in
Figure~\ref{images_sub}, these results point 
to a smaller ionized volume of He than H. This,
in turn, points to a relatively cool star(s) responsible for the 
ionizing radiation field (Osterbrock 1989)\markcite{ob89}. 

According to Osterbrock (1989)\markcite{ob89}, 
equal volumes of ionized He and H do not
occur until the effective temperature of the ionizing star is \teff \apge 
40,000~K.
For spherical volumes and an ionization 
bounded H~II region, Osterbrock (1989)\markcite{ob89} has plotted the 
ratio of radii between the He and H volumes vs. \teff \
(his Figure 2.5). 
For the blister geometry (taken here to be a
one dimensional plane--parallel slab), 
the ratio of volumes changes only linearly with 
depth into the cloud since the area of the two volumes is the same but 
with the ionized H zone extending 
deeper into the cloud. 
One can consider applying Osterbrock's plot to
a blister seen edge on where the hydrogen and helium ionized 
zones have a ratio of 
volumes $= d_{\rm He}/d_{\rm H}$, where $d$ is the depth of the 
blister ionized volume into the cloud. 
For the present case, an upper limit to \teff \ is suggested by considering
the ratio of thicknesses of the He~I and Br$\gamma$ emission regions
(Figure~\ref{images_sub} and \S 3.1). Then, $d_{\rm He}/d_{\rm H}$ $=$ 
6.3$''$~/~8.5$''$ $=$ 0.74. The ratio of radii for an equivalent 
ratio of spherical
volumes is $(d_{\rm He}/d_{\rm H})^{1/3}$ $=$ 0.9 and Osterbrock's plot
indicates \teff \ $=$ 38,800~K. This is an approximate upper limit because the 
blister may be seen in projection, rather than strictly edge on.
An approximate 
lower limit comes from considering the ratio of radii which enclose
the total flux in He~I and Br$\gamma$ (Figure~\ref{faper}). Taking these as
the radii in Osterbrock's plot, we have $r_{\rm He}/r_{\rm H}$ $=$
15$''$~/~35$''$ $=$ 0.43, and  \teff \ $=$ 35,000 K.
More detailed calculations (Shields 1993\markcite{js93}) for a wide range of
parameters, including the effects of dust in the nebula, 
also 
suggest that \teff \ $=$ 40,000 K represents the minimum temperature for the
ionizing source at which the He$^{+}$ zone is equal in extent to the H$^{+}$
zone.

The rough estimate above is supported by analysis of the observed lines through
comparison to ionization models.
We have computed blister models using the ionization code Cloudy
(Ferland 1996)\markcite{gf96}; visit
http://server1.pa.uky.edu:80/$\sim$gary/cloudy/ . A blister is an ionized
slab of material on the face of a molecular cloud.
For Cloudy, the geometry (i.e. the analysis) is always one dimensional 
(radial). 
Therefore, an idealized blister geometry is a plane parallel 
slab made by considering a shell 
illuminated by an ionizing source from a large distance on one face. 
This idealized geometry is only an approximation to the real case
represented by Figures~\ref{images} and \ref{images_sub}. We will compare
average observed quantities to the one dimensional models.

We consider only uniform,
non--clumpy slabs. An important input parameter
is the ionizing photon density at the cloud (slab) face (set by the
ionization parameter for a given gas density; see below). 
A given ionization parameter can describe more than one model:
a more luminous 
source which is further away should produce the same ionization as a less
luminous source closer to the cloud.
The ionizing source 
may be covered by gas (closed geometry), 
or it may stand off the slab in an open geometry. 
The line luminosity will be approximately proportional to the solid angle
of gas subtended by the ionizing source. This is 
the covering factor. Only second order
effects due to the radiation transfer of the diffuse radiation field will
occur between open and closed geometries (Ferland 1996\markcite{gf96}), so 
the choice should not be important in considering line ratio discussed below.
We have used an open geometry. Finally, our models assume the H~II region is
ionization bounded; all the photons emitted by the ionizing source toward the 
cloud are absorbed (up to a scaling factor given by the covering factor
which cancels in the line ratio).

The models use Kurucz (1991)\markcite{k91} 
continua and the ``\_ism'' abundances (including the effects of dust,
and with N$_{\rm He}$/N$_{\rm H}$ $=$ 0.1). The detailed ``\_ism'' abundance
mixture is given by Ferland (1996)\markcite{gf96}.
Several tests with varying log(g) showed little effect on the resulting
ratios of interest, so we chose log(g) $=$ 5 to avoid having to interpolate
in the high \teff \ models. We adopted a hydrogen density of 2000 cm$^{-3}$
(Liszt 1992\markcite{l92}; Mehringer et al. 1998\markcite{mea98}).

\begin{figure}
\plotfiddle{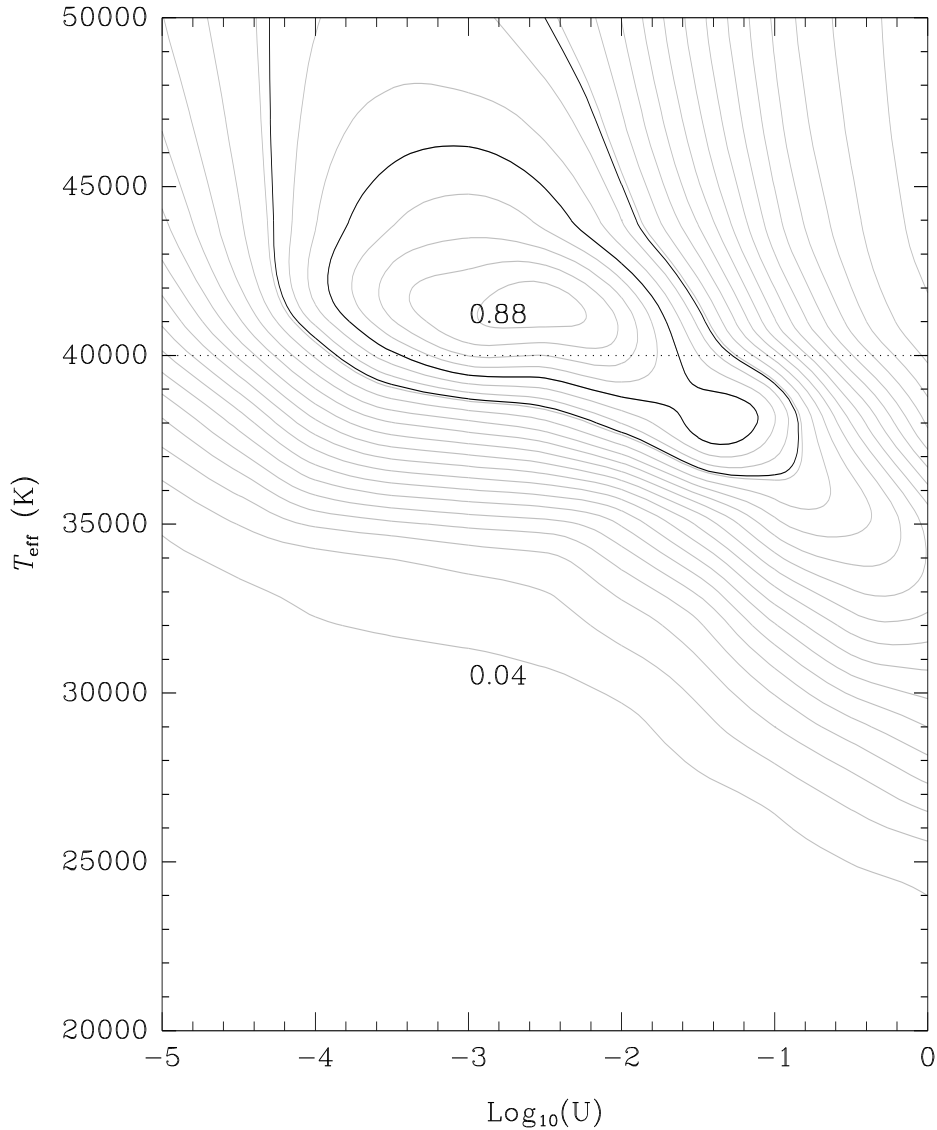}{5 in}{0}{75}{75}{-220}{-80}
\figcaption[]{He~I 2.06 \mic \ to Br$\gamma$ line ratio contour plot. 
The plot was
generated from a grid of Cloudy ionization models with varying 
stellar \teff \ and ionization parameter, U; see text.
The contour interval is 0.04. Basic contours
are plotted as grey, while the contours corresponding to $A_K = 1, 3$ 
mag (0.61 and
0.72, respectively) are plotted as black. 
The {\it dotted} line is an approximate upper limit on \teff \ based on the
larger extent of the Br$\gamma$ emission relative to He~I; see text.
This plot shows differences from earlier versions of Cloudy; see \S 3.2.
\label{206}
}
\end{figure}

\begin{figure}
\plotfiddle{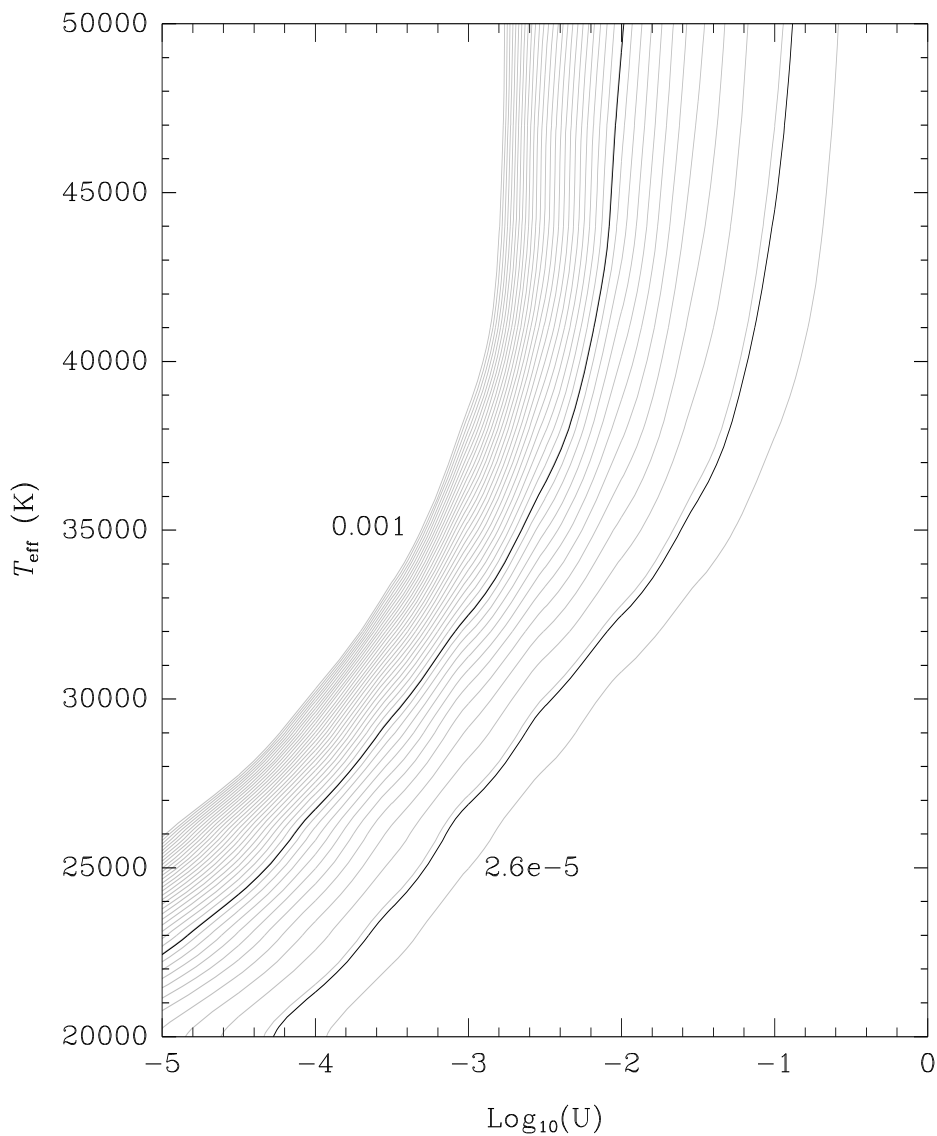}{5 in}{0}{75}{75}{-220}{-80}
\figcaption[]{Br$\gamma$ / far infrared dust emission contour plot. 
The far infrared flux is from Odenwald \& Fazio (1984).
The plot was generated from a grid of Cloudy ionization models 
with varying stellar \teff \ and ionization parameter, U; see text.
The contour interval is 2.6$\times$10$^{-5}$, and contours above
0.001 are not plotted.
Basic contours are plotted as grey, while the contours corresponding to 
$A_K = 1, 3$ mag (4.7e-5 and 3.1e-4, respectively) are plotted as black. 
\label{grain}
}
\end{figure}

We computed grids of ionization models parameterized by the \teff \ of the
ionizing source and the ionization parameter, U $\equiv$ $\phi / n(H)c$, where
$\phi$ is the number flux of ionizing photons (cm$^{-2} s^{-1}$), $n(H)$ is
the hydrogen density, and $c$ is the speed of light. Thus, for $n, c$ in 
cm$^{-3}$ and cm~s$^{-1}$, respectively, 
U is a non--dimensional parameter which sets the ionizing
photon number density at the incident cloud face.
From the model grids we
produced contour plots of the He~I 2.06 \mic \ to Br$\gamma$ line ratio
(Figure~\ref{206})
and the ratio of Br$\gamma$ to far infrared flux 
(Figure~\ref{grain}). For the latter ratio, we adopt the measurement
of Odenwald \& Fazio (1984 \markcite{of84}) for the far infrared flux
(40--250 \mic). Odenwald and Fazio report a size of 1.1$'$ for the Sgr~D
far infrared source. This agrees with the 35$''$ radius we attribute to the
H~II region; therefore, we compare the fluxes directly. The Br$\gamma$
and He~I lines are relatively close in wavelength, so the line ratio
does not change a great deal with $A_K$. 
We use the total He~I flux in a 15$''$ radius aperture and the corresponding
flux from the same size aperture for Br$\gamma$. This gives a ratio of
He~I to Br$\gamma$ of 0.56 $\pm$ 0.04, 
which ranges from 0.72 to 0.61 when corrected
for reddening (1 \aple $A_K$ \aple 3 mag based on the results of \S~3.1.1). 
The ratio of Br$\gamma$ to far infrared flux varies from $4.7 \times 10^{-5}$
to $3.1 \times 10^{-4}$ for the same range in $A_K$. 
For $A_K$ $=$ 1.5 mag, as indicated by the results of \S 3.1.1, the corrected
ratio of He~I to Br$\gamma$ is 0.64, and the Br$\gamma$ to far infrared flux
ratio is $7.6 \times 10^{-5}$. 

\begin{figure}
\plotfiddle{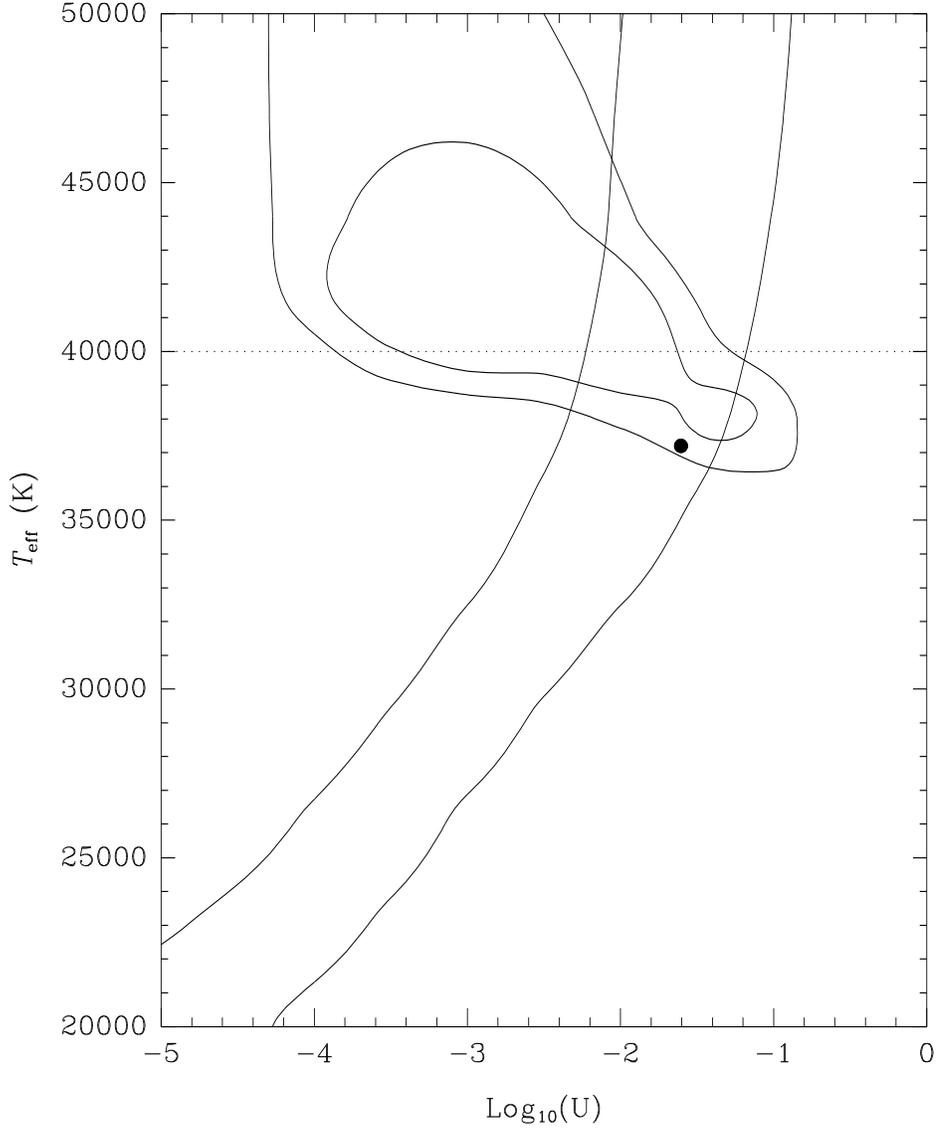}{5 in}{0}{75}{75}{-220}{-80}
\figcaption[]{
Overlap region for the allowed contours of Figures~\ref{206} and \ref{grain}.
The {\it dotted} line represents an approximate 
upper limit based upon the larger
extent to the Br$\gamma$ emission than for the He~I emission; see text.
For the indicated region outlined by the contours and {\it dotted} line,
\teff \ of the ionizing radiation field is 
approximately 37,000~K to 40,000~K. For a specific case of $A_K$ $=$ 1.5 mag
(as suggested by measurements of $A_K$ for stars projected against the 
dark cloud),
the {\it black dot} at 37,200 K represents the intersection of contours for the
He~I to Br$\gamma$ ratio and the Br$\gamma$ to far infrared flux ratio.
\label{over}
}
\end{figure}

The \teff \ of the ionizing radiation field can be estimated 
from Figures~\ref{206}
and \ref{grain} by overlaying the contours which bracket the observations
(for the range of $A_K$ taken above).
An overlay of the allowed contours from
Figures~\ref{206} and \ref{grain} is shown in Figure~\ref{over}.
We find 36,500~K $\leq$ \teff \ $\leq$
40,000~K, in agreement with our rough estimate based on the extent
of the He~I and Br$\gamma$ emission zones. We have taken 40,000 K as an
approximate
upper limit to the \teff \ of the ionizing radiation field based on the
fact that the observed 
Br$\gamma$ emission is more extended than the He~I emission 
(see the discussion above).
For the specific case of $A_K$ $=$ 1.5 mag, the appropriate contours (0.64
and $7.6 \times 10^{-5}$ in Figures~\ref{206} and \ref{grain}, respectively)
intersect in Figure~\ref{over} at \teff $=$ 37,200 K (in the allowed region). 
The resulting electron temperatures 
for the  models in the overlap region are between $\sim$ $7,000 - 8,000$~K. 

The upper contours in Figure~\ref{206} are closed. This indicates the ratio
peaks and then begins to fall again at higher \teff. This double
valued behavior
is discussed in detail by Shields (1993\markcite{js93}) and is due to
the complex behavior of the He~I line radiative transfer
for higher \teff \ (\apge 40,000 K) of the ionizing source. The physical 
arguments laid out by Shields remain valid, but the detailed 
model results (also obtained with Cloudy)
for higher  \teff \ have changed slightly since his work due to 
changes implemented in the Cloudy code. 
The behavior of the He~I to Br$\gamma$
line ratio below \teff \ \aple 40,000 K is very similar for the 
present code and the the version used by 
Shields (1993\markcite{js93}).
The modifications to the current version of Cloudy
are discussed by Ferland (1999\markcite{gf99}).

For an ionization bounded 
H~II region and case B conditions, we can estimate the 
number of ionizing photons required to produce the Br$\gamma$ emission
assuming $A_K$ $=$ 1.5 mag as indicated by the measurments of stars
near the H~II region (\S 3.1.1). Taking the distance to the 
H~II region as 6.0 kpc $\pm ^{1.5} _{2.5}$ kpc (\S~3.1.2), 
gives 48.55 $\leq$ log($Q_{\circ}$ s$^{-1}$) $\leq$ 48.99.
Liszt (1992\markcite{l92}) and Mehringer et al. (1998\markcite{mea98})
report 
48.31 $\leq$ log($Q_{\circ}$ s$^{-1}$) $\leq$ 48.75 and 
48.65 $\leq$ log($Q_{\circ}$ s$^{-1}$) $\leq$ 49.09,
respectively (corrected to the present distance estimate).
Therefore, the Br$\gamma$ 
results are in good agreement with the radio data for the same 
assumed distance and the estimated $A_K$.
Alternately,
the extinction at 2.17 \mic \ can be estimated by using 
the observed radio flux to predict the expected intrinsic Br$\gamma$ flux and
then comparing this latter number to the observed Br$\gamma$ flux. 
We estimate the average extinction at 2.17 \mic \ by using the predicted Lyman
continuum photon luminosity 
from the radio data (this depends on the integrated radio flux)
to estimate the Br$\gamma$ luminosity (e.g., Osterbrock
1989\markcite{ob89} equation 5$-$23).
For case B conditions and $T_{\rm e}$ $=$ 8000 K, the ratio of ionizing
photons to Br$\gamma$ luminosity (in cgs) is 7.2$\times$~10$^{13}$ using 
the calculations by Hummer \& Storey (1987)\markcite{hs87} to determine
the Br$\gamma$ effective recombination coefficient. 
The observed Br$\gamma$ flux is taken from Figure~\ref{faper} at 35$''$.
We find $A_{2.17}$
$=$ 1.01 mag and 1.86 mag using the number of ionizing photons from 
Liszt (1992\markcite{l92}) and Mehringer et al. (1998\markcite{mea98}),
respectively. This is 
consistent with the independent measurements of nearby stars
found in \S 3.1.1.

\subsection{Location of the H~II Region}

Lis (1991)\markcite{l91} and Mehringer et al. (1998\markcite{mea98}) 
provided strong evidence 
that Sgr~D is not associated with the GC nuclear disk GMCs:
the molecular emission and 
absorption line widths were significantly narrower than for
other GMCs in the nuclear disk. Placing Sgr~D on either side of the 
GC is more difficult. 
Lis (1991\markcite{l91}) argues that the
distance is not less than about 3.5 kpc based on a calculated cloud
size and observed angular size.
A nearby distance to Sgr~D would have important consequences for
the star formation rate in the GC (see Lis 1991\markcite{l91}
for a discussion). 
Lis also suggested Sgr~D could be a feature of the expanding molecular
ring, still in the GC region. This fits with the observed CS emission
from the cloud associated
with the H~II region 
which is at negative velocity (Lis 1991\markcite{l91}) and perhaps the 
recombination line velocity which is also negative (Liszt 1992\markcite{l92}).
The latter is less clear since the recombination line is for the {\it 
extended} H~II region.
Mehringer et al. (1998\markcite{meal98}) 
argue that the Sgr~D H~II region could be significantly farther than 
the GC based on H$_{2}$CO absorption by intervening clouds of the {\it 
extended} H~II region continuum. This absorption has a large width and 
positive mean velocity which they 
associate with a GC cloud. Thus, the Sgr D H~II region must be beyond the GC 
to be seen in absorption. The present narrow--band images appear to
rule this out for the compact H~II region and dark cloud
associated with it.

Our results require a location at least on the near side of the GC. This
is essentially because the stellar density in front of the dark cloud
is less than half what it is off the cloud (\S 3.1.2). 
The stellar
distribution peaks so sharply at the GC that the point at which half the stars
are observed along a line of sight will be at or before the GC.
For the dark cloud on the narrow--band images to occult as many stars as 
it does, it must be on the near side of the GC. A model which accounts for
all the observations is, then, the suggestion of Lis (1991\markcite{l91}) that
the dark cloud associated with the {\it compact}
H~II region is part of the expanding molecular ring. It is in the Galactic 
center region, but on the near side. 

\section{SUMMARY}

We have presented 2 \mic \ narrow--band images of the Sgr~D H~II region
including continuum subtracted He~I 2.06 \mic \ and Br$\gamma$ emission lines.
The images reveal the compact H~II region previously described at
far infrared and longer wavelengths. Our images 
confirm the earlier suggestion by Odenwald (1989) that the H~II region is a 
blister on the edge of a dark cloud. The region in projection against the 
dark cloud has far fewer stars compared to the other areas of
the frame which are crowded with the dense inner Galaxy stellar field. 
This in itself strongly argues that the cloud must be on the near side
of the GC.
Our star count models and \ak \ determinations are consistent with a
location of the H~II region between $\sim$ 3.9 and 7.9 kpc (for a sun--to--GC
distance of 8 kpc). A scenario which is consistent with our results and 
previous molecular line measurements is the suggestion by 
Lis (1991\markcite{l91}) that the cloud associated with the compact
H~II region is located in the expanding molecular ring.

Our analysis of
the He~I 2.06 \mic \ emission relative to Br$\gamma$ indicates the 
ionizing radiation field arises from a source(s) 
with 36,500~K \aple \teff \ \aple
40,000~K (for a range of \ak \ of 1 to 3 mag). The upper limit is set by the
fact that the Br$\gamma$ emission is more extended than the He~I emission.
For an \ak \ corresponding to the closest stars in projection 
to the H~II region
($\sim$ 1.5 mag) the derived \teff \ is 37,200~K.
The total number of ionizing photons required to produce
the de-reddened 
Br$\gamma$ luminosity (for an ionization bounded H~II region  with 
case B conditions) is in good agreement with previous estimates from
radio continuum observations for the same assumed distance to the H~II
region.

We thank J. Shields and G. Ferland for useful discussions and 
extensive help with the ionization models. We thank K. Sellgren for useful
discussions and a helpful reading of the manuscript.
We appreciate the comments and suggestions of an anonymous referee which 
greatly improved our paper.
Support for this work was provided by NASA through grant number
HF 01067.01 -- 94A from the Space Telescope Science Institute, which is
operated by the Association of Universities for Research in Astronomy,
Inc., under NASA contract NAS5--26555. A.D. acknowledges financial support 
received from PRONEX/FINEP.
Thanks also to B. Dylan for the new album.

\newpage


\newpage


\end{document}